\documentclass[conference]{IEEEtran}
\IEEEoverridecommandlockouts
\usepackage{cite}
\usepackage{multirow}
\usepackage{amsmath,amssymb,amsfonts}
\usepackage{algorithmic}
\usepackage{graphicx}
\usepackage{textcomp}
\usepackage{xcolor}
\usepackage{float}
\def\BibTeX{{\rm B\kern-.05em{\sc i\kern-.025em b}\kern-.08em
    T\kern-.1667em\lower.7ex\hbox{E}\kern-.125emX}}

\begin{document}

\title{An Attention-based Long Short-Term Memory Framework for Detection of Bitcoin Scams




\thanks{\IEEEauthorrefmark{1} Corresponding author.

This work is partly supported by Zhuhai Basic and Applied Basic Research Foundation Grant ZH22017003200018PWC, and partly supported by the Guangdong Provincial Key Laboratory of 
Interdisciplinary Research and Application for Data Science, BNU-HKBU United International 
College, Project code 2022B1212010006 and in part by Guangdong Higher Education Upgrading
Plan (2021-2025) UIC  R0400001-22.}
}

\author{\IEEEauthorblockN{Puyang Zhao\IEEEauthorrefmark{1}, Wei Tian\IEEEauthorrefmark{1},  Lefu Xiao\IEEEauthorrefmark{1}, Xinhui Liu\IEEEauthorrefmark{1}, Jingjin Wu\IEEEauthorrefmark{1}\IEEEauthorrefmark{2}}
\IEEEauthorblockA{\IEEEauthorrefmark{1}Department of Statistics and Data Science, BNU-HKBU United International College, Zhuhai, Guangdong, P. R. China \\
\IEEEauthorrefmark{2}Guangdong Provincial Key Laboratory of Interdisciplinary Research and Application for Data
Science}
Email: puyangzhao.27@gmail.com; s230202702@mail.uic.edu.cn; p930005056@mail.uic.edu.cn; \\
xinhui\_liu@outlook.com; jj.wu@ieee.org\\}

\IEEEoverridecommandlockouts
\IEEEpubid{\makebox[\columnwidth]{978-1-6654-9144-0/22/\$31.00~\copyright2022 IEEE \hfill} \hspace{\columnsep}\makebox[\columnwidth]{ }}

\maketitle
\begin{abstract}

Bitcoin is the most common cryptocurrency involved in cyber scams. Cybercriminals often utilize pseudonymity and privacy protection mechanism associated with Bitcoin transactions to make their scams virtually untraceable. The Ponzi scheme has attracted particularly significant attention among the Bitcoin fraudulent activities. This paper considers a multi-class classification problem to determine whether a transaction is involved in Ponzi schemes or other cyber scams, or is a non-scam transaction. We design a specifically designed crawler to collect data and propose a novel Attention-based Long Short-Term Memory (A-LSTM) 
method for the classification problem. The experimental results show that the proposed model has better efficiency and accuracy than existing approaches, including Random Forest, Extra Trees, Gradient Boosting, and classical LSTM. With correctly identified scam features, our proposed A-LSTM achieves an F1-score over 82$\%$ for the original data and outperforms the existing approaches.

\end{abstract}


\section{Introduction}

Bitcoin is the first decentralized cryptocurrency. 
As of October 2021, Bitcoin had a market share of around $45\%$, being the highest among all cryptocurrencies~\cite{coindesk2018}, and is expected to continue dominating the crypto market in the foreseeable future.
This paper will study techniques to detect cyber-crime activities conducted by Bitcoin. 

There are multiple forms of cybercrime involving Bitcoin transactions, such as Ponzi schemes, cryptojacking, and e-mail frauds. Among these, Ponzi schemes represent one of the most prevalent types of cybercrime. Statistics show that almost $\$$7 billion was generated in cryptocurrency revenue by Ponzi schemes in 2019, nearly twice the amount generated by all other cyber fraud categories combined in 2020~\cite{grauer20212021}. 

A general trend is that more and more investors are becoming victims of cyber scams involving cryptos due to inadequacies ineffective intervention and prevention measures. 
Thus, one of the essential steps is to detect cyber scams in their early stages to ensure the proper functioning of the cyber society. In this paper, we classify all Bitcoin transactions into three categories: 1) transactions involved in a Ponzi scheme, 2) transactions involved in other types of scams, or 3) normal non-scam transactions, for the sake of preventing the scams in advance or detecting them in the early stage of the fraud. 

In this paper, we develop a framework that can accurately detect Ponzi schemes and other scams conducted by Bitcoin transactions with a novel deep learning method called attention-based Long Short-Term Memory (A-LSTM).  
\textbf{The main contributions are summarized as follows.} 
\begin{itemize}
    \item We design a crawler which can automatically crawl information of Bitcoin transactions that potentially involve scams from known Bitcoin addresses, such that we can obtain the firsthand information. The crawler automatically parses websites based on a dictionary that contains Ponzi-related words like ``Ponzi", ``profit", ``HYI", ``multiplier", ``investment", ``MLM". With the crawler, we manage to collect a number of Bitcoin addresses that initiated transactions, and then build a dataset considerably larger than those used in similar existing studies. 
    \item From the transaction information, we study the features that distinguish normal transactions from those involving cyber scams. We identify the five most influential features in Bitcoin scams detection, providing insights into the detection of such scams. They are (i) active days; (ii) number of outs; (iii) input number; (iv) the total number of BTC spent; (v) number of addresses received. The features would be explained in detail later. 
    \item We adopt the A-LSTM mechanism that suits the features of our constructed dataset to classify the transactions in our framework. We compare the performance of our proposed A-LSTM approach with four popular supervised learning models, namely Random Forest~\cite{liaw2002classification}, Extra Trees~\cite{geurts2006extremely}, Gradient Boosting~\cite{friedman2001greedy} and classical LSTM~\cite{hochreiter1997long}. 
    We also integrate resampling methods with each of these methods, aiming to solve the imbalance problem in the dataset. We demonstrate that, while resampling is a traditional method for solving the imbalance problem, it is not applicable to the A-LSTM model. This is because the resampling method would introduce a large amount of noise into A-LSTM. 
    On the other hand, A-LSTM without resampling gives even better results than other methods with resampling.

\end{itemize}

\begin{figure}[h]
     \centering
     \includegraphics[scale=0.35]{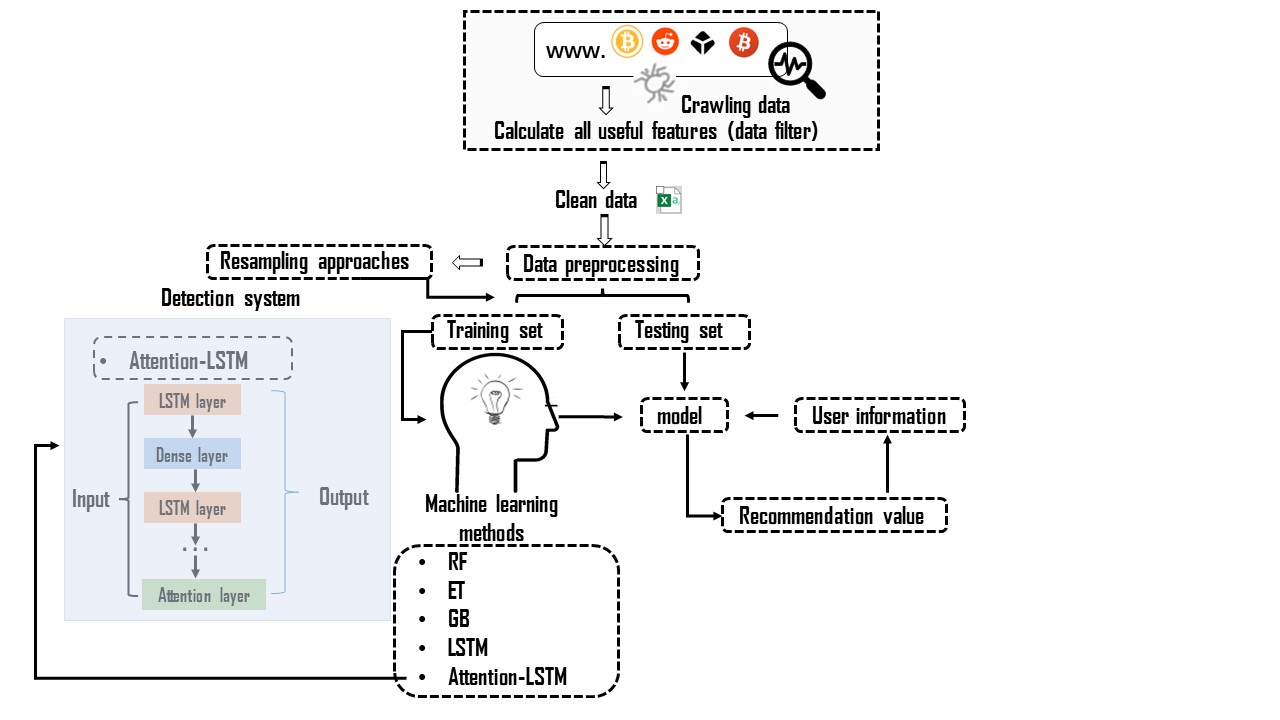}
     \caption{\textit{Workflow of our Bitcoin scam detection framework}}
     \label{Workflow}
 \end{figure}

Fig.~\ref{Workflow} presents an overview of our proposed framework. 
The rest of this paper is organized as follows. Section II summarizes the existing related works. Section III describes our methodologies for identifying and collecting addresses of Ponzi schemes, and for constructing a data set of Ponzi scheme related features. Section IV describes the steps of the proposed classification approaches in detail. Section V compares the effectiveness of strategies in terms of correctly classifying transactions. Finally, Section VI concludes the paper and provides some potential future research directions.

\section{Related Works}

Since Bitcoin scammers frequently change their IP address to evade from cyber regulators, it is quite difficult to gather sufficient data to perform meaningful analysis. Most existing Bitcoin scam studies collected the Bitcoin addresses involved through manual or semi-automated searches\cite{Brenig2015EconomicAO}. Nevertheless, it should be noted that these methods would fail when the fraudulent addresses are not disclosed, for example, in private communication or when a transaction is conducted on the deep or dark web. Therefore, manual or semi-automated collection method is relatively inefficient when the majority of scams are carried out using hidden addresses. Another consequence of this fact is that, although the number of scams and relevant studies 
is rapidly increasing, few public datasets are available for further analysis. For example, in Bartoletti et al.~\cite{bartoletti2018data}, a small and imbalanced dataset consisting of 32 Ponzi scam cases and 6,400 normal cases was considered. Some more recent studies (e.g.,~\cite{fan2021spsd}) combined the datasets of several previous studies aiming at making meaningful conclusions with sufficient data. However, a downside of this approach is that the timeliness of the dataset is not preserved, and the features of scams from different periods may not be the same.
Therefore, when dealing with these cases, we believe that it is optimal to use tools that automatically search~\cite{bartoletti2018data} the Bitcoin blockchain for suspicious behavior and identify new addresses associated with fraudulent activities within a rather short period of time.

In terms of classifiers, the most popular ones used in existing studies include RIPPER~\cite{cohen1995fast} and Random Forest (RF). RF was the best classifier, having obtained a F1 score of $78.7\%$ for Ponzi schemes according to~\cite{bhattacharyya2011data}.
While more data mining methods were employed to detect Ponzi schemes in recent studies, the highest F1 scores among them have not exceeded that in~\cite{bhattacharyya2011data}. A potential reason is that other Bitcoin scams, such as sextortion, blackmail scam, and pyramid schemes~\cite{buil2021offending}, are also increasing. These scams share certain similar features as Ponzi schemes. Existing data mining-based classifiers for this purpose are all binary, which means that they only distinguish whether a transaction is involved in a Ponzi scheme or not, and thus may cause plenty of false positives for detecting Ponzi schemes~\cite{fan2021spsd}.

To the best of our knowledge, our proposed method is the first to add attention mechanisms to identify Bitcoin scams. Detailed explanations for the new mechanism will be provided in later sections. 

\section{Data Collection}

Our work is arranged as follows. First, we collect Bitcoin addresses that investors could use to send money to scam operators. The next step is to create a set of features relevant to scam classification, compute their values on our addresses, and then train classifiers based on this dataset. After that, we formalize a detection model for Bitcoin scams as a multi-class classification problem, where the task was to distinguish Ponzi schemes, other scams, and non-scam transactions. 


\subsection{Collection of Bitcoin addresses}

We first search Reddit and Bitcointalk.org, the two largest Bitcoin communities, to collect addresses that conducted Bitcoin transactions. We investigate advertisements related to High-Yield Investment Programs (HYIPs), which are investment schemes that promise extraordinarily high returns, up to $100\%$, as a majority of HYIPs are Ponzi schemes. Then, we search each web page through the address on the advertisement to find the Bitcoin address where the funds were deposited. In rare cases, these advertisements will explicitly state where the funds are deposited. Therefore, we can also obtain the address by visiting the website hosting the program. 

Moreover, we thoroughly study ``The 2021 Crypto Crime Report"~\cite{grauer20212021} to identify  Bitcoin addresses involved in Ponzi schemes and other scams, and collect Bitcoin addresses from other existing relevant papers with publicly available data. In addition, we find Bitcoin addresses related to cyber scams reported on the Bitcoin Abuse Database website. Some addresses found in previous steps are duplicated and discarded. Overall, we identify 160 deposit addresses involved in Ponzi schemes and 442 involved in other scams. A small proportion of the addresses are shown in Table~\ref{t1}. 

\begin{table}[htbp]
\caption{A Collection of Bitcoin addresses}
\begin{center}
\begin{tabular}{|c|c|}
\hline
\textbf{Types}&\textbf{Bitcoin address} \\
\hline
Ponzi scheme&19g9exzmtJ2sQbBBB3x2PiY9pReVCm8HqA\\
Ponzi scheme&1DXxLzocfWXTHVYv4MTai4LLtXZgcJDknZ\\
Other scam&12t9YDPgwueZ9NyMgw519p7AA8isjr6SMw\\
Other scam&1CamL5swGf1sVCkzuya6TyE1KXAT3xmdxZ\\
Non-scam&1AfSRAGjxx5azGtCgboxnAP1F6pB95VALW\\
Non-scam&1FTjwjX5M7CHESQCJ3p1T1GNe1R17V8o9t\\
$\cdots$&$\cdots$\\
\hline
\end{tabular}
\label{t1}
\end{center}
\end{table}

\subsection{Data Extraction and Dataset Construction}
Open-source tools make extracting data easier due to their low complexity. Using the API of Bitcoin, we extract transaction data from different Bitcoin addresses. Then, we access data on each transaction by Blockchain APIs for traders. However, the amount of data on the website makes manual extraction computationally prohibitive. To accomplish this, we develop a web crawler to parse data corresponding to different Bitcoin addresses, and sort collected data for the convenience of further analysis.

A publicly available crawler, based on the Bitcoinabuse API, is able to obtain many Bitcoin addresses that were involved in different types of scams. However, an address testing indicates that many duplicate addresses and false addresses (that is, addresses that actually do not have transaction records or invalid addresses) are present in the original dataset. Therefore, it is necessary to revise the original crawler such that it can also handle data cleaning and pre-processing tasks. 

Specifically, after getting the address from the crawler, we first check if the address has relevant transaction data. If a certain transaction contains no data or is a duplicate message, it will be discarded. In addition, we normalize and standardize the characteristics when necessary. As an example, we uniformly convert the life cycle and delay characteristics into days since the original records in seconds are sometimes too large for processing, which significantly impacts the prediction results. 



As we mentioned earlier, \textcolor{black}{each collected transaction maybe 1) a transaction involved in a Ponzi scheme, 2) a transaction involved in a type of scam other than Ponzi Scheme, or 3) a normal transaction. We divided the data into a training set and a test set with a ratio of 80:20.
The number of transactions in the collected dataset belonging to each category is shown in Table \ref{t3class}.}

\begin{table}[htbp]
\caption{Breakdown of Instances in the dataset}
\begin{center}
\begin{tabular}{|c|c|c|}
\hline
Type  & No. of instances \\
\hline
Normal Instances&  3008\\
\hline
Ponzi Schemes&  285\\
\hline
Other Scams&  3526\\
\hline
\end{tabular}
\label{t3class}
\end{center}
\end{table}





\subsection{Features extraction}

Now, we will introduce the features used for our classification of Bitcoin transactions. Based on the study \cite{bartoletti2018data}, we define relevant features as follows:

\begin{itemize}
\item \textbf{Lifetime.} The number of days between the first and last transaction to/from the address.
\item \textbf{Active day.} The total number of days with at least one transaction.
\item \textbf{Most Active day.} The number of transactions in a day with the most transactions.
\item \textbf{Number In/Out.} The number of transactions received or spent.
\item \textbf{In vs Out.} The ratio of the number of transactions received to the number of transactions spent.
\item \textbf{Number of address received/spent.} The number of addresses that sent/received Bitcoin transactions.
\item \textbf{Median/Mean delay.} The mean/median of the interval between receiving a transaction and sending a transaction.
\item \textbf{Minimum/Maximum delay.} The minimum/maximum value of the interval between receiving a transaction and sending a transaction.
\item \textbf{Total received/spent BTC.} The total value of BTC received by/sent from this address.
\item \textbf{48 diff.} The largest income and expenditure difference for this Bitcoin address within 48 hours.
\end{itemize}

From the dataset as well as the intuitive knowledge on crypto transactions, there exist serious proportion imbalances across the three categories, with the vast majority of transactions being normal transactions. Such imbalance may lead to inaccuracy in prediction results and has to be addressed properly, either by traditional resampling methods or novel approaches. 

\section{Multi-class Classification}

We now describe our approaches for the Bitcoin scam detection, a multi-class classification problem. We will particularly describe our proposed A-LSTM method that can address imbalances among classes, and introduce performance measures and validation protocols. Our last step is to determine the elements in our collection that are the most relevant to detect scams.

\subsection{Oversampling and undersampling techniques}
Similar to the situation in real life, our dataset is imbalanced, where the scams only account for a tiny proportion. Therefore, we consider traditional resampling methods, which aim to obtain a balanced sample distribution by changing the original unbalanced sample set and learning a suitable model. 



We will consider the following variants of oversampling and undersampling methods, and combine them with different data mining techniques (to be described in the next subsections). Classical oversampling methods include ROS, synthetic minority oversampling technique (SMOTE)~\cite{chawla2002smote}, and adaptive synthetic (ADASYN)~\cite{he2008adasyn} sampling approach. 
We will also include two most recently proposed resampling approaches: 1) SMOTE and Edited Nearest Neighbor Method (SMOTE-ENN), a hybrid strategy that combines oversampling and undersampling techniques~\cite{sahare2012review}, and 2) Tomek Links~\cite{elhassan2016classification}, another undersampling method modified from the Condensed Nearest Neighbors. 

\subsection{Traditional Data Mining Methods}

We consider the following popular existing methods as benchmarks in this paper.

\subsubsection{\textbf{Random Forest (RF)}}
RF is a classifier containing multiple decision trees, and its output category is determined by each tree of each output category~\cite{liaw2002classification}. Given training dataset $D = \{x_i,y_i\}_i^n,x_i$ with n observations and m features $\in \mathcal{R}^p$,$y_i\in\{0,1,2\}$. RF bootstraps $D^*_j$ $j\in\{1,A\}$ from $D$ and selects $F = \sqrt{m}$ features. Then RF will find out the best split features from the subsets.



\subsubsection{\textbf{Extra Trees}}
Extra Trees,
also called Extremely Randomized Trees~\cite{geurts2006extremely}, is a variant of RF. For the training set of each decision tree, extra trees generally do not use random sampling, that is, each decision tree uses the original training set. After the division features are selected, extra trees randomly select a feature value to further divide the decision tree. 
In some cases, the generalization ability of extra trees is better than RF, and ET's can be computationally faster. However, ET performs a little worse when there are a lot of noisy features.

\subsubsection{\textbf{Gradient Boosting (GB)}}

GB can be regarded as gradient descent over the function space~\cite{friedman2002stochastic}. In GB, each iteration generates a weak learner that fits the gradient of the loss function with respect to the previous cumulative model, and then adds this weak learner to the cumulative model to gradually reduce the loss of the cumulative model. In our classification problem, we consider the logistic loss function updated based on its negative gradient. 


 \begin{figure}[h]
     \centering
     \includegraphics[scale=0.35]{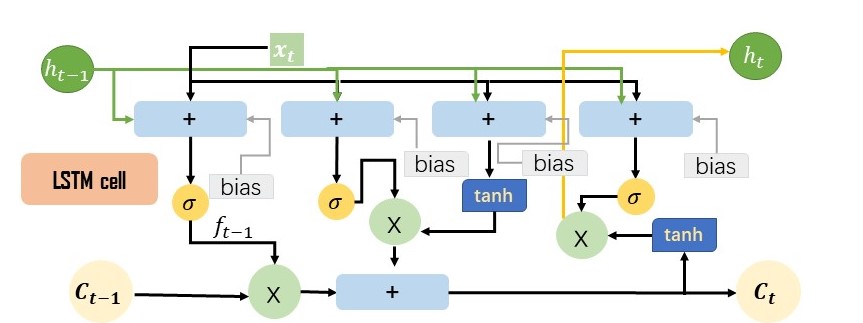}
     \caption{\textit{The structure of an LSTM cell}}
     \label{lstm_1}
 \end{figure}
 
 \begin{figure}[h]
     \centering
     \includegraphics[scale=0.4]{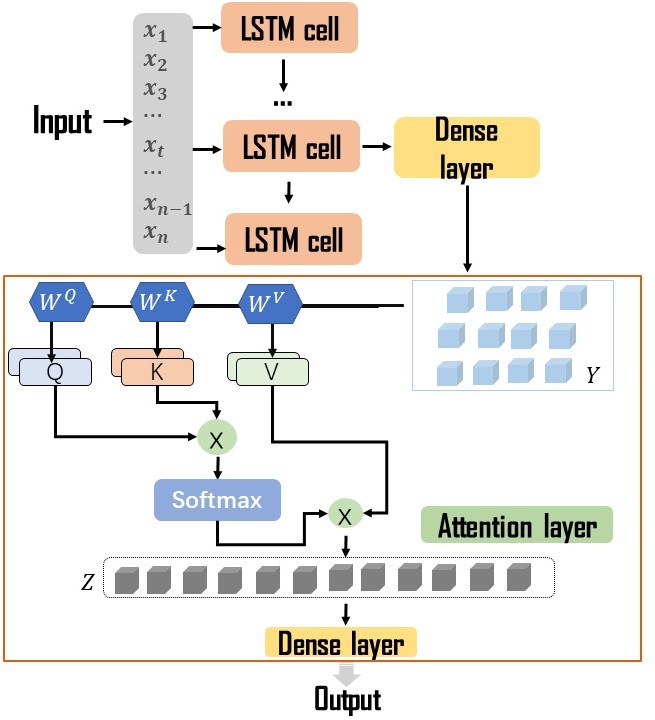}
     \caption{\textit The structure of the proposed A-LSTM model}
     \label{structure}
 \end{figure}
\subsection{Our Proposal}

We now describe our proposed new neural network model for classifying the Bitcoin dataset, A-LSTM.

Long Short Term Memory is based on a recurrent neural network designed for processing sequential data~\cite{hochreiter1997long}. It evolved from recurrent neural networks (RNN) and solves the gradient vanishing problem of RNN in capturing long-range information. LSTM has been demonstrated to achieve good performance in time series data processing, such as in the field of cryptocurrency. Fig.~\ref{lstm_1} shows an LSTM cell with input, forget, and output gates. 
The weighted sum of the outputs 
from all those three gates will be the LSTM unit output. Here, $h_t$, the output of the LSTM 
at time $t$, can be calculated by
\begin{equation}
h_t=o_t*\tanh{(i_t*c_t'+f_t c_{t-1})}\label{eqh}
\end{equation}
where $o_t,i_t,f_t$ are outputs from the output, input and forget gates at 
time $t$, $c_t^{'}(. )$ and $c_t(. )$ represent the candidate and final states of the memory cell at time $t$. 

Our proposal adds an \textit{attention mechanism} to the LSTM technique to better capture global information. The structure of our proposed A-LSTM is shown in Fig.~\ref{structure}, where an attention layer and a dense layer are added to further process the output produced by LSTM. 

The rationale for adding the attention mechanism is that, in Bitcoin trading, the importance of various fraudulent information varies, some information is redundant while others may be critical. In fraud detection, we need to focus on the key features first and discard the redundant features. Therefore, valid information from different addresses can be used to build a fraud detection model. The attention mechanism, being able to capture the valuable information in the training set, can then improve the accuracy of detection. 

In addition, the long-term dependency of the original LSTM would cause the decoder to send an incorrect representation of the current step dependency. This may lead to poor output from the decoder as the original LSTM does not address the dependency issue. In contrast, our proposed attention mechanism is capable of handling the dependence among the LSTM outputs. This can also be reflected in the attention weights matrices $Q$ (query), $K$ (key value), and $V$ (value), computed from self-attention, as shown in Fig.~\ref{structure}. Thus, by better representing important parts of the input sequence in the decoder input better, the self-attention mechanism can facilitate better decoder output.

The attention mechanism focuses on computing the ``weight" of each value in the set of values. We obtain matrices $Q$, $K$, and $V$ by packing the embeddings into a matrix $Y$ and multiplying it by three weighting matrices ($W_Q, W_K, W_V$). Then, we will get the output $Z$ of the attention layer. Specifically,
\begin{equation}
Y * W_Q = Q
\end{equation}
\begin{equation}
Y * W_K = K
\end{equation}
\begin{equation}
Y * W_V=V
\end{equation}
\begin{equation}
Z=\text{softmax}(Q×K^T)V\label{eqa}
\end{equation}


\subsection{Performance measures}
A multi-class problem can generally be converted to a two-class problem for simplicity. In this regard, the label of the minority class is positive, and the label of the majority class is negative. Ponzi schemes or other scams are considered the minority class, respectively, while normal instances are the majority class.

We consider the following performance measures that are commonly used in classification problems. 
\begin{itemize}
\item \textit{Precision} is the ratio of correctly predicted positive observations to the total predicted positive observations. High precision indicates that our model can detect most of the frauds. 

\item \textit{Recall} is the ratio of correctly predicted positive observations to all observations. A high Recall value indicates that the classifier can correctly identify the entries belonging to the minority class. 

\item \textit{Accuracy} is the ratio of the correctly predicted observations to the total observations. It is considered the most intuitive measures of performance. 

\item The \textit{F1-score} combines the precision and recall of a classifier into a single metric, which is a comprehensive measure of a classification model's accuracy. The formula of F1-score is
\end{itemize}

\begin{equation}
\text{F1-Score} = \frac{2\times (\text{Recall}\times \text{Precision})}{\text{Recall} + \text{Precision}}\label{eq3}
\end{equation}


We consider the change in the \textit{Gini index} caused by branching a certain feature vector $\mathbf{x}_j$, denoted by $\text{VIM}_j^{(\text{Gini})}$, to represent the relative importance of feature $j$. With $M$ feature vectors $\mathbf{x}_1,\mathbf{x}_2, \cdots, \mathbf{x}_M$, the Gini index of $\mathbf{x}_m$ ($m \in \{1,2,\cdots,M\}$) is 
\begin{equation}
\text{GI}_m = \sum_{k=1}^{|K|}\sum_{k'\neq k}P_{mk}P_{mk'}=1 - \sum_{k=1}^{|K|}P_{mk}^2\label{eqGNI}
\end{equation}
where $K$ is the number of classes, and $P_{mk}$ represents the proportion of class $k$ in node $m$. 

The relative importance of feature $j$ at node $m$ is then the change in Gini index before and after the branch of node $m$, namely 
$\text{VIM}_{j}^{(\text{Gini})} = \Delta \text{ GI}_m$, where a higher value of $\text{VIM}_{j}^{(\text{Gini})}$ indicates that feature $j$ is more important.

\section{Results}

\begin{figure}[ht]
     \centering
     \includegraphics[scale=0.33]{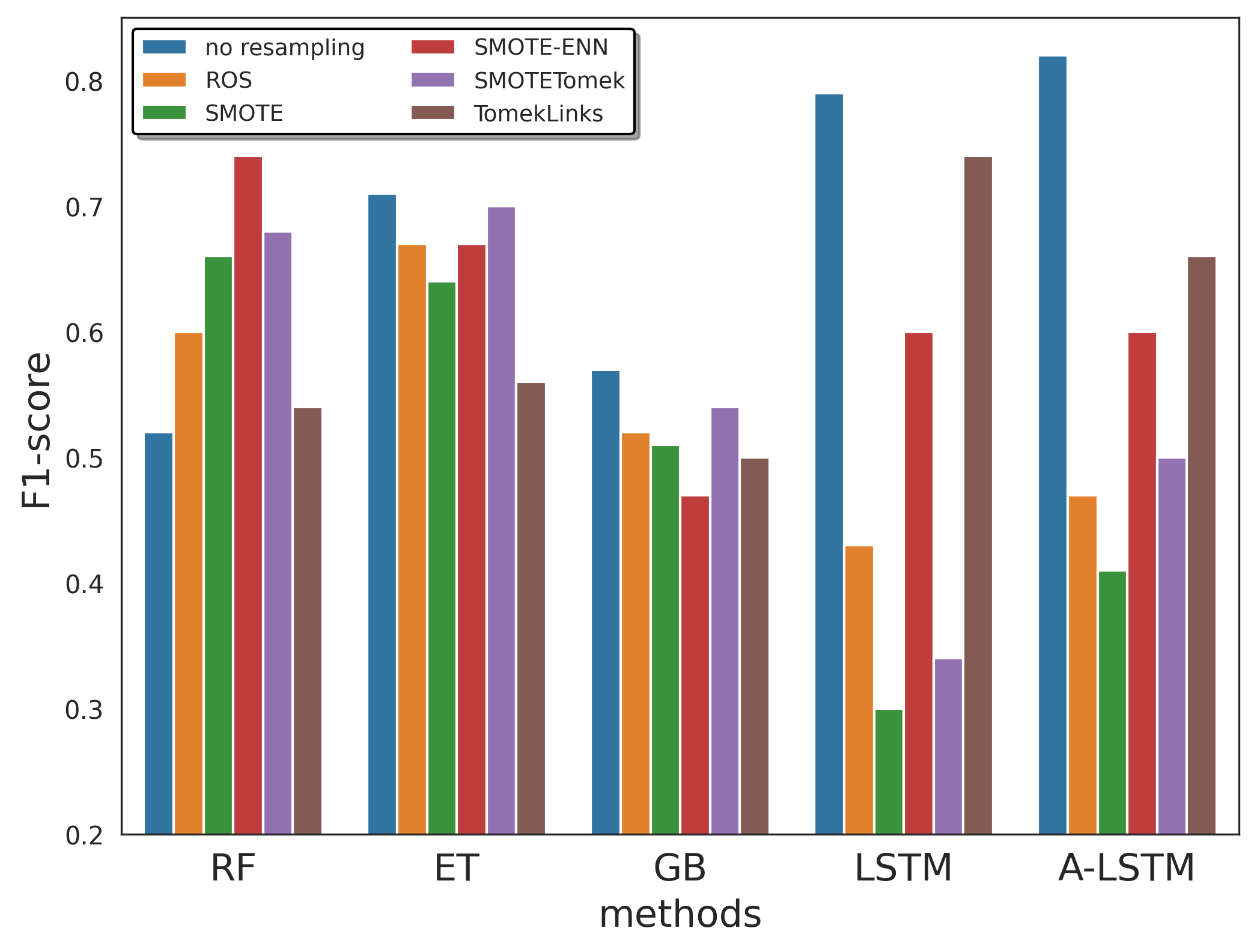}
     \caption{\textit{F1-scores of different combinations of classification approaches and resampling methods.}}
     \label{r}
 \end{figure}

In this section, we present and compare the classification results by different combinations of classifiers and resampling techniques. Accuracy, Precision and Recall of selected combinations are shown in Table~\ref{t5}, while F1-scores  are shown in Fig.~\ref{r}.

\begin{figure}[H]
    \centering
    \includegraphics[scale=0.1]{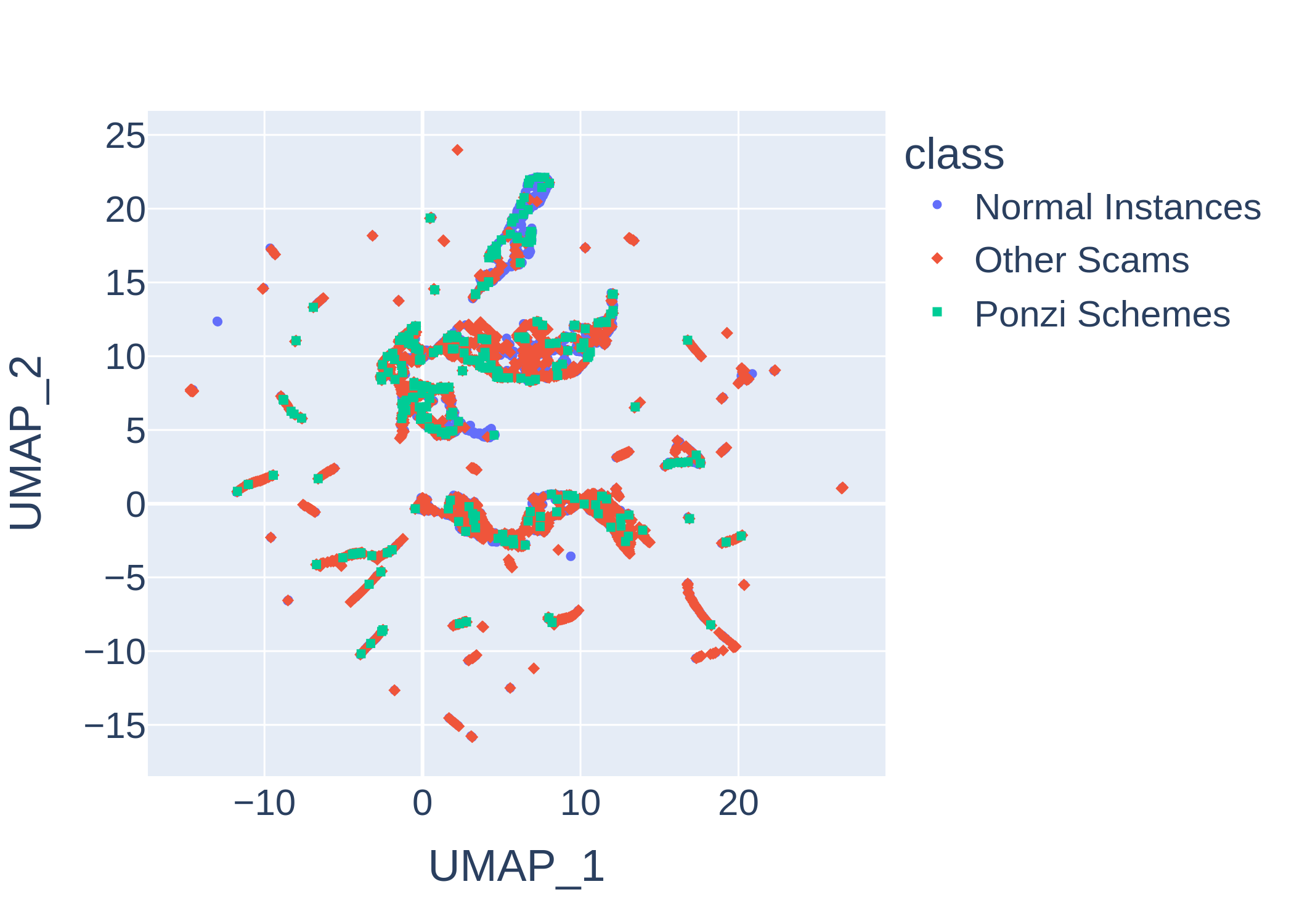}
    \caption{\textit{UMAP for our Bitcoin dataset}}
    \label{f13}
\end{figure}

Among the conventional data mining techniques without resampling, LSTM achieves the highest F1-score of $79\%$, followed by RF's $74\%$. Both are considerably better than ET and GB. The poor performances of ET and GB can be partly explained by the UMAP in Fig.~\ref{f13}, which shows that our dataset is hard to be separated after dimensionality reduction. On the other hand, our proposed A-LSTM gives the best results in terms of all measurements, particularly with an F1-score as high as $82\%$. 
The resampling approaches can generally improve the performance of some other classification methods but fail for A-LSTM. 
For some of the methods, resampling does not improve but deteriorate the classficiation accuracy. This is mainly because that resampling may increase the bias, especially when the noise in the training dataset insignificant as in our scenario. For A-LSTM in particular, resampling methods can make deep learning methods like A-LSTM severely overfitting and lead to poor results. Moreover, the training process of deep learning can over-amplify the imbalance issue if resampling is adopted. 
Among resampling approaches, TomekLinks has the most robust performance, mainly because it can overcome the interference present in all other methods. Also, TomekLinks is an undersampling method, which does not add much noise impact, but makes the dataset smaller and leads to undertraining. Therefore, it also does not improve the A-LSTM results.

To summarize, while selected resampling approaches can achieve certain improvements after integrating with resampling approaches, none of the combinations could outperform our proposed A-LSTM. Therefore, the A-LSTM can be regarded as a new and promising approach for detecting Bitcoin scams.

\begin{figure}[htbp]
\centering
    \includegraphics[scale=0.28]{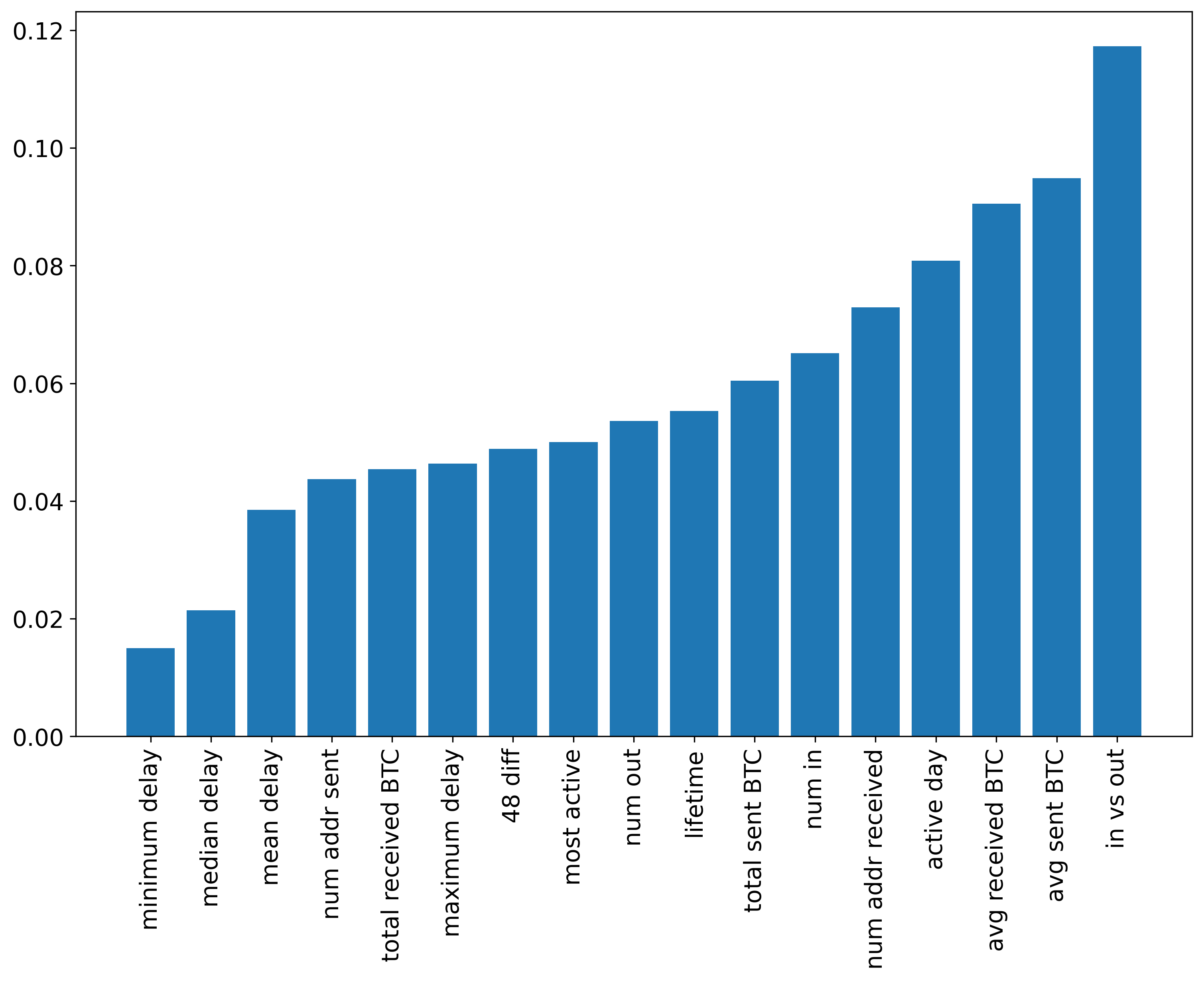}
    \caption{\textit{Relative importance of features in classification}}
    \label{f_8}
\end{figure}

Finally, we rank the relative importance of features for the classification of scams. The results are presented in Fig.~\ref{f_8}, with the features sorted by the change in the Gini index in ascending order. 
The five most discriminating features, in ascending order of relevance, are (i) in vs out; (ii) avg received BTC; (iii) avg sent BTC; (iv) total spent BTC; (v) total received BTC. While the rank is somewhat counter-intuitive (for example, lifetime might be important from a straightforward perspective), this result can provide useful guidelines for the data collection process of future similar research, such that one may focus on the features that are important. 

\section{Concluding Remarks}

In this work, we proposed a framework to effectively collect Bitcoin transaction data, and classify whether a transaction is a scam. We particularly focused on the Ponzi scheme, a classic scam disguised as a "high-yield" investment plan, among other cyber scams. For the data collection task, our specifically designed crawler managed to collect a considerably larger dataset than existing studies on similar topics. For classification, we proposed the A-LSTM approach, based on the deep learning mechanism, and demonstrated that it outperformed existing popular approaches with or without the resampling technique in terms of the F1-score. 
However, we use a relatively small dataset in this paper due to difficulties in the data crawling process. We plan to find new methods to expand the data in the future, such as Generative Adversarial Network to 
further improve our dataset. We also analysed the relative importance of features to classify transactions involved in Ponzi schemes or other scams, providing valuable insight for future research in related fields.



\begin{table}[htbp]
\caption{Performance comparison in terms of accuracy, precision and recall of different selected approaches}
\begin{center}
\begin{tabular}{|c|c|c|c|c|}
\hline
Method  & Accuracy & Precision & Recall \\\hline
{RF \& Nonsampling}  & 0.60     &0.74      &0.60\\ \hline
{RF \& ROS} & 0.82     &0.83       &0.82\\\hline
{ET \& Nonsampling} & 0.73     &0.76      &0.73\\ \hline
{ET \& SMOTETomek}  & 0.72     &0.76       &0.72\\
\hline
{GB \& Nonsampling}  & 0.57     &0.57      &0.57\\ \hline
{GB \& ROS}  & 0.60     &0.72       &0.60\\\hline
{LSTM \& Nonsampling}  & 0.80    &0.80      &0.80\\ \hline
{LSTM \& SMOTETomek}  &   0.78     &0.26       &0.05\\
\hline
{A-LSTM \& Nonsampling}&0.83& 0.85& 0.80\\\hline
{A-LSTM \& ROS} &0.38& 0.74&  0.38 \\\hline
{A-LSTM \& SMOTE} &0.36&  0.63 & 0.36\\\hline
{A-LSTM \& SMOTE-ENN}&0.62&  0.75 & 0.62\\\hline
{A-LSTM \& SMOTETomek}  &0.40& 0.79 & 0.40\\\hline
{A-LSTM \& TomekLinks} &0.83& 0.68& 0.68\\\hline

\end{tabular}
\label{t5}
\end{center}
\end{table}

\bibliographystyle{IEEEtran}
\bibliography{IEEEexample}
\vspace{12pt}
\end{document}